# Asymmetric Electron Energy Loss in Drift-Current Biased Graphene

Filipa R. Prudêncio[1,2]* and Mário G. Silveirinha[1]

[1] *University of Lisbon – Instituto Superior Técnico and Instituto de Telecomunicações, Avenida Rovisco Pais 1, 1049-001 Lisboa, Portugal*

[2]*Instituto Universitário de Lisboa (ISCTE-IUL), Lisboa, Portugal*

## Abstract

The electric drift current bias was recently introduced as a new paradigm to break the Lorentz reciprocity in graphene. Here, we study the impact of the nonreciprocal response in the energy extracted from a beam of swift charges travelling in the vicinity a graphene sheet with drifting electrons. It is demonstrated that the drift bias leads to an asymmetric electron-energy-loss spectrum that depends on the sign of the charges velocity. It is found that when the drift and electron beam velocities have comparable values but opposite signs, the energy loss is boosted resulting in a noncontact friction-type effect. In contrast, when the drift and electron beam velocities have the same sign the energy loss is negligible. Furthermore, it is shown that different theoretical models of the drift-biased graphene conductivity yield distinct peaks for the energy-loss spectrum, and thereby electron beam spectroscopy can be used to test the validity of the different theories.

---

* Corresponding author: filipa.prudencio@lx.it.pt

# I. Introduction

Nonreciprocal photonic platforms have been extensively investigated in the literature and offer unique opportunities to control electromagnetic wave propagation [1-9]. In particular, gyrotropic material structures can enable one-way light flows due to the nonreciprocal response of the involved materials [1-8]. Such solutions require an external magnetic bias to break the Lorentz reciprocity, which is often impractical [10-11]. Systems relying on nonlinear effects [12-13], opto-mechanical interactions [14-16], transistor-loaded metamaterials [17-18], or spacetime modulations [19-22], offer alternative solutions to produce strong nonreciprocal responses without static magnetic fields.

The nonreciprocity of gyrotropic media is rooted in the magnetic field bias that forces the electrons to undergo cyclotron orbits in the bulk region. The circular motion of the charged particles creates a drag-type effect when the electrons interact with a time varying field, which is at the origin of the Faraday rotation in bulk gyrotropic media. A simple way to picture the wave propagation in a magnetically biased material is to imagine that the wave goes through a centrifuge, e.g., rotating drum, and is dragged by the rotation of the walls. The static magnetic field endows the material with an angular momentum.

A different paradigm for a magnetic-free nonreciprocal response was recently studied by different authors [6], [23-29], and relies on a graphene sheet with drifting electrons. In this case, a static voltage generator biases the electrons with a "linear momentum", which is manifested in the form of DC current; thus, a drift-current biased graphene may behave similarly to a moving medium [6, 27-30]. In particular, the drifting electrons may drag the graphene plasmons along the direction of motion, analogous to the Fizeau effect [31]. For large drift velocities comparable to the Fermi velocity $v_\text{F}$, the drag effect can be so strong that the plasmon propagation becomes unidirectional and is largely insensitive to the backscattering from obstacles or imperfections [6]. Furthermore, a drift-current biased graphene sheet is an active system because the electrons can give away their kinetic energy in the form of electromagnetic radiation [27, 29].

The electrons in graphene are described by a massless Dirac equation and behave like relativistic Dirac Fermions with a linear dispersion [32]. The peculiar electronic properties of graphene lead to rather unique features of the optical response, e.g., to a strongly spatially dispersive conductivity with a square-root singularity in the momentum space [32]. A particularly attractive feature of graphene is the ultra-subwavelength nature of plasmon excitations with relatively large propagation lengths. Graphene plasmonics is expected to play an important role in the development of THz photonics [33], and may have important applications in waveguiding, biological sensing, spectroscopy, and others.

Electronic energy loss spectroscopy (EELS) is a powerful method to characterize electronic band structures, plasmons, and the response of structured materials [34-37]. It relies on the analysis of either the electron energy losses or of the emitted radiation when a beam of swift electrons travels close by (or through) some target. In particular, low-energy electron microscopy uses electrons with 1-100 eV, and enables a sub-eV energy resolution (on the order of 20meV or less using monochromated electron beams [40]) which is fine enough to characterize low-energy excitations, such as acoustic plasmons and excitons in the infrared range [36-40]. The amount of energy loss measured via EELS depends on phonon excitations, plasmons excitations and Cherenkov radiation [41]. Recently, the Cherenkov effect was investigated in plasmonic platforms and in metamaterials with a plasmonic-type hyperbolic response [42]-[49]. In particular, in Ref. [49] we found out that in topological nonreciprocal plasmonic systems the spectrum of the emitted Cherenkov radiation can be strongly dependent on the direction of motion of the electron beam.

Motivated by these developments, here we investigate how the drift-current bias of graphene may influence the power extracted from a beam of swift electrons that travel parallel to the graphene sheet. We find that the energy loss is typically boosted when the electron beam moves in the direction *opposite* to the drifting electrons. In contrast, the energy loss is negligible when the electron beam and the drifting electrons move along the same direction with similar velocities. Thus, the drift-current creates a strongly asymmetric noncontact friction effect. Furthermore, we compare the energy loss predicted by different conductivity models of the drift-current biased graphene. Our calculations suggest that EELS measurements can be useful to test the validity of the available theoretical models and to characterize the dispersion of short-wavelength plasmons.

## II. Radiation Mechanisms

In our theoretical analysis the electron beam is modeled as a pencil beam (Fig. 1) described by the current density $\mathbf{j}(x,z,t) = -e n_y v_0 \delta(z-z_0) \delta(x-v_0 t) \hat{\mathbf{x}}$. Here, $v_0$ is the velocity of moving charges, $-e$ is the electron charge, $n_y$ is the number of charges per unit of length along the $y$-direction, and $z_0 \equiv d$ is the height of the beam relative to the graphene sheet. For simplicity, the beamwidth, measured along the $z$-direction, is taken infinitesimally small. The problem is effectively two-dimensional because $\mathbf{j}$ is independent of $y$. It is assumed that the graphene sheet lies on the top of material substrate (e.g., $SiO_2$) with a dielectric constant $\varepsilon_s = 4$ and is biased with drifting electrons with a velocity $v_{\text{drift}}$. The drift current is induced by a DC voltage generator (not shown in Fig. 1).

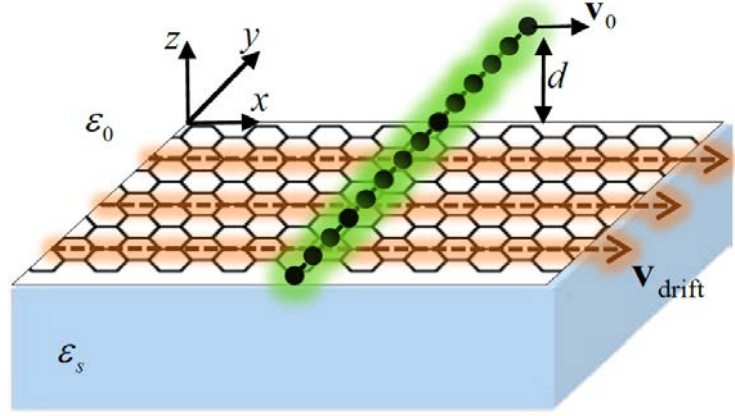

**Fig. 1** A pencil beam of electrons with velocity $v_0$ moves nearby a graphene sheet biased with drifting electrons with a velocity $v_{drift}$. The graphene sheet stands on the top of a dielectric (SiO$_2$) substrate.

Following Ref. [49], the instantaneous magnetic field excited by the electron beam is of the form $\mathbf{H}(x,z,t) = H_y \hat{\mathbf{y}}$ with

$$H_y(x,z,t) = \frac{1}{2} n_y e \, \text{sgn}(v_0) \omega_F \, h_y(x,z,t)$$
$$h_y(x,z,t) = \frac{1}{2\pi} \int_{-\infty}^{+\infty} h_y(x,z,\omega) e^{-i\omega t} \frac{d\omega}{\omega_F} \quad , \tag{1}$$

where $\omega_F$ is a normalization constant with unities of frequency so that $h_y(x,z,t)$ is dimensionless. The normalization constant is taken identical to $\omega_F = \mu_c / \hbar$, where $\mu_c$ is the chemical potential of the graphene sheet and $\hbar$ is the reduced Planck constant. The fields have a travelling-wave type variation in space-time, such that $H_y(x,z,t) = H_y(x-v_0 t, z, t=0)$. The function $h_y(x,z,\omega)$ determines the spectrum of the emitted magnetic field. In the vacuum region ($z > 0$), it can be written as

$$h_y(x,z,\omega) = \left[ \text{sgn}(-z+z_0) e^{-\gamma_0 |z-z_0|} + R e^{-\gamma_0 (z+z_0)} \right] e^{ik_x x}, \quad k_x = \omega / v_0. \tag{2}$$

Here, $\gamma_0 = \sqrt{k_x^2 - \omega^2/c^2}$ is the z-propagation constant in free-space and $R = R(\omega, k_x)$ is the (magnetic field) reflection coefficient of the graphene sheet for plane wave incidence. In the previous formulas, the wave number $k_x$ must satisfy the selection rule $k_x = \omega/v_0$, and thereby is indirectly determined by the energy of the swift electrons. The electric field in the air region is such that $\partial_t \mathbf{E} = \varepsilon_0^{-1}(\nabla \times \mathbf{H} - \mathbf{j})$.

By enforcing the continuity of the tangential component of the electric field $E_x\big|_{z=0^+} - E_x\big|_{z=0^-} = 0$ and the impedance boundary condition $H_y\big|_{z=0^+} - H_y\big|_{z=0^-} = -\sigma_{g,\mathrm{drift}} E_x$, it can be shown that [27]

$$R(\omega, k_x) = \frac{i\omega\varepsilon_0(\varepsilon_s\gamma_0 - \gamma_s) - \sigma_{g,\mathrm{drift}}\gamma_s\gamma_0}{i\omega\varepsilon_0(\varepsilon_s\gamma_0 + \gamma_s) - \sigma_{g,\mathrm{drift}}\gamma_s\gamma_0}, \tag{3}$$

where $\gamma_s = \sqrt{k_x^2 - \varepsilon_s \omega^2/c^2}$ is the $z$-propagation constant in the substrate and $\sigma_{g,\mathrm{drift}}$ is the graphene conductivity with the drifting electrons.

One can find in the literature different theoretical models for the conductivity of graphene with a drift current bias [23-29]. In Refs. [27, 29] it was shown using distinct approaches that when the electron scattering is mainly determined by electron-electron interactions, the effect of the electric bias is ruled by a Galilean Doppler-shift ($\omega \to \tilde{\omega} = \omega - k_x v_{\mathrm{drift}}$ and $k_x \to \tilde{k}_x = k_x$), such that the conductivity is given by:

$$\sigma_{g,\mathrm{drift}}(\omega, k_x) = \frac{\omega}{\tilde{\omega}} \sigma_g(\tilde{\omega}, k_x), \qquad \tilde{\omega} = \omega - k_x v_{\mathrm{drift}}. \tag{4}$$

In the above, $\sigma_g(\omega)$ is the bare graphene conductivity (with no drift), $v_{\mathrm{drift}}$ is the drift velocity, $\tilde{\omega}$ is the Doppler shifted frequency, $\omega$ is the oscillation frequency, and $k_x$ is the propagation constant along de $x$-direction. The time variation $e^{-i\omega t}$ is implicit. Alternative models for the graphene conductivity were developed by other authors [23-26] relying on a shifted Fermi distribution. With a single band approximation (with the interband transitions discarded), the conductivity obtained with the shifted Fermi distribution is ruled by a relativistic-type Doppler shift in the zero temperature limit [23, 29]:

$$\sigma_{g,\text{drift}}^R(\omega, k_x) = \frac{\omega}{\tilde{\omega}} \sigma_g(\tilde{\omega}, \tilde{k}_x), \qquad \tilde{\omega} = \gamma_g(\omega - k_x v_{\text{drift}}), \qquad \tilde{k}_x = \gamma_g(k_x - \omega v_{\text{drift}}/v_F^2) \quad (5)$$

where $\gamma_g = 1/\sqrt{1 - v_{\text{drift}}^2/v_F^2}$ is the graphene Lorentz factor and $v_F$ is the Fermi velocity.

As discussed in Ref. [29], the shifted Fermi distribution determines a distribution of the electronic kinetic momentum, whereas the relevant distribution for the computation of the graphene conductivity with the Lindhard formula is a distribution of canonical (Bloch wave vector) momentum, which is unaffected by the drift current. Thus, in our understanding, Eq. (4) should model better the effect of the drift-current bias than Eq. (5). Intuitively, when the electron-electron collisions predominate the electron gas is forced to move as a whole (with the drift velocity) similar to a moving medium, and thereby its conductivity is expected to be described by a standard Galilean Doppler shift transformation [27, 29]. We demonstrate later in the article that the energy loss spectrum predicted by Eqs. (4) and (5) may differ substantially, and thereby that EELS measurements may be useful to experimentally test the validity of the theoretical models.

The bare (no-drift) graphene conductivity $\sigma_g(\omega, k_x)$ is calculated analytically using [50, 51]

$$\sigma_g(\omega, k_x) = -i\omega \chi_\Gamma(\omega, q), \qquad (6a)$$

$$\chi_\Gamma(\omega, q) = \left(1 + \frac{i\Gamma}{\omega}\right) \frac{\chi(\omega + i\Gamma, q)}{1 + \frac{i\Gamma}{\omega} \frac{\chi(\omega + i\Gamma, q)}{\chi(0, q)}}, \qquad (6b)$$

$$\chi(\omega, q) = \frac{e^2}{4\pi\hbar}\left[\frac{8k_F}{q^2 v_F} + \frac{G(\Delta_-) - G(\Delta_+) + i\pi}{\text{sq}(\omega - v_F q)\text{sq}(\omega + v_F q)}\right], \qquad (6c)$$

where $\Gamma$ is the scattering rate, $k_F = \mu_c/(\hbar v_F)$ is the Fermi wave number, $\mu_c$ is the chemical potential, $\Delta_\pm = (\hbar\omega \pm 2\mu_c)/(\hbar v_F q)$ and $q = \sqrt{k_x^2}$. We rewrote the formula of Refs. [50-51] in such a way that it is an analytical function of $k_x = k_x' + ik_x''$ when $k_x$ is near the real-axis and $\omega$ is real-valued. The functions $G(z)$ and $\mathrm{sq}(z)$ in Eq. (6c) are defined in the Appendix. The formalism assumes that $k_B T \ll \mu_c$. Throughout the article, we use $\mu_c = 0.1\,\mathrm{eV}$ and a scattering rate equal to $\Gamma = 1/(0.35\,\mathrm{ps})$.

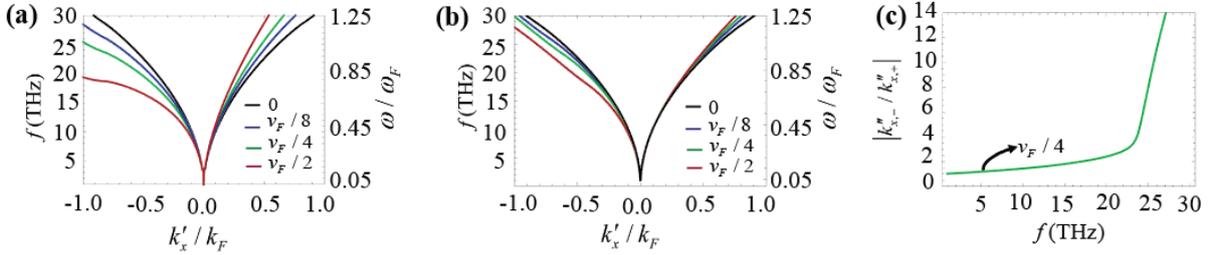

**Fig. 2 (a)** Dispersion diagram of the graphene plasmons [Fig.1] calculated with the Galilean Doppler-shift model of graphene [Eq. (4)] for different drift velocities indicated in the insets; **(b)** similar to (a) but calculated with relativistic Doppler shift model [Eq. (5)]. **(c)** Ratio of the attenuation constants for counter-propagating plasmons, for $v_\mathrm{drift} = v_F/4$, obtained with the Galilean Doppler-shift model. The complex propagation constant is $k_x = k_x' + ik_x''$.

To begin with, we calculate the dispersion of the graphene plasmons for different drift velocities. The dispersion equation is determined by the poles of the reflection coefficient: $i\omega\varepsilon_0(\varepsilon_s\gamma_0 + \gamma_s) - \sigma_{g,\mathrm{drift}}\gamma_s\gamma_0 = 0$ [see Eq. (3)]. For simplicity, we use a quasi-static approximation ($\gamma_0 \approx \gamma_s \approx q$), which yields $\dfrac{i\omega\varepsilon_0}{\sigma_{g,\mathrm{drift}}}(\varepsilon_s + 1) - q = 0$. The dispersion diagram is found by solving this equation with respect to $k_x = k_x' + ik_x''$ as a function of a real-valued $\omega$. The dispersion for the Galilean Doppler-shift model of the conductivity is shown in Fig. 2a for drift velocities $v_\mathrm{drift}$ on the order of $v_F = c/300$ [52-53]. As seen, the drift bias creates an asymmetry between the +x and –x directions such that the dispersion diagram bends upwards (downwards) for waves that co-propagate (counter-propagate) with the drifting electrons, creating conditions for regimes of unidirectional propagation. Figure 2c shows the ratio of the imaginary parts ($k_+''$ and $k_-''$) of the complex propagation constant for counter-propagating plasmons and $v_{\text{drift}} = v_F / 4$. Note that $k_+''$ and $k_-''$ are the attenuation constants associated with waves propagating along the +x and –x-directions, respectively. As seen, the waves propagating in the direction opposite to the drifting electrons (associated with $k_-''$), are much more attenuated than the waves propagating along the drift direction. Thus, a drift-biased graphene sheet can effectively behave as a one-atom thick optical isolator [6, 28]. For comparison, we show in Fig. 2b the dispersion diagram obtained with the relativistic Doppler shift model. The results are qualitatively similar to those depicted in Fig. 2a, but the effect of the drift on the dispersion of the graphene plasmons is dramatically weaker in a relativistic model. In the rest of article, except if explicitly mentioned otherwise, all the calculations will be based on the Galilean Doppler-shift model of graphene [Eq. (4)].

Let us now suppose that an electron beam interacts with the drift-current biased graphene. The moving electrons interact resonantly with the graphene plasmons that satisfy the selection rule $k_x = \omega / v_0$ [36, 50] (it is underlined that $R$ in Eq. (2) is evaluated with $k_x = \omega / v_0$). Thus, the plasmons that can be excited by moving charges with $\pm v_0$ velocity can be found by intersecting the lines $k_x = \omega / v_0$ and $k_x = \omega / -v_0$ with the plasmons dispersion diagram, respectively. This is illustrated in Fig. 3a for the case of $v_{\text{drift}} = v_F / 4$ and an electron beam with velocity $\pm v_0 = \pm 2 v_F$; this velocity to corresponds to an electron energy on the order of 11 eV. As seen, the propagation constant and oscillation frequency of the excited plasmons are strongly dependent on the sign of the velocity of moving charges $v_0$ due to the asymmetry of the dispersion diagram. Consistent with this property, the absolute value of the reflection coefficient $R(\omega, k_x)|_{k_x=\omega/v_0}$ is peaked at a frequency that depends on the sign of $v_0$ (Fig. 3b). For example, for $v_0 = \pm 2v_F$, $|R|$ is peaked near 13.5 THz and 37.1 THz, respectively, which are the frequencies determined by the intersection of $k_x = \pm\omega/v_0$ with the plasmons dispersion diagram.

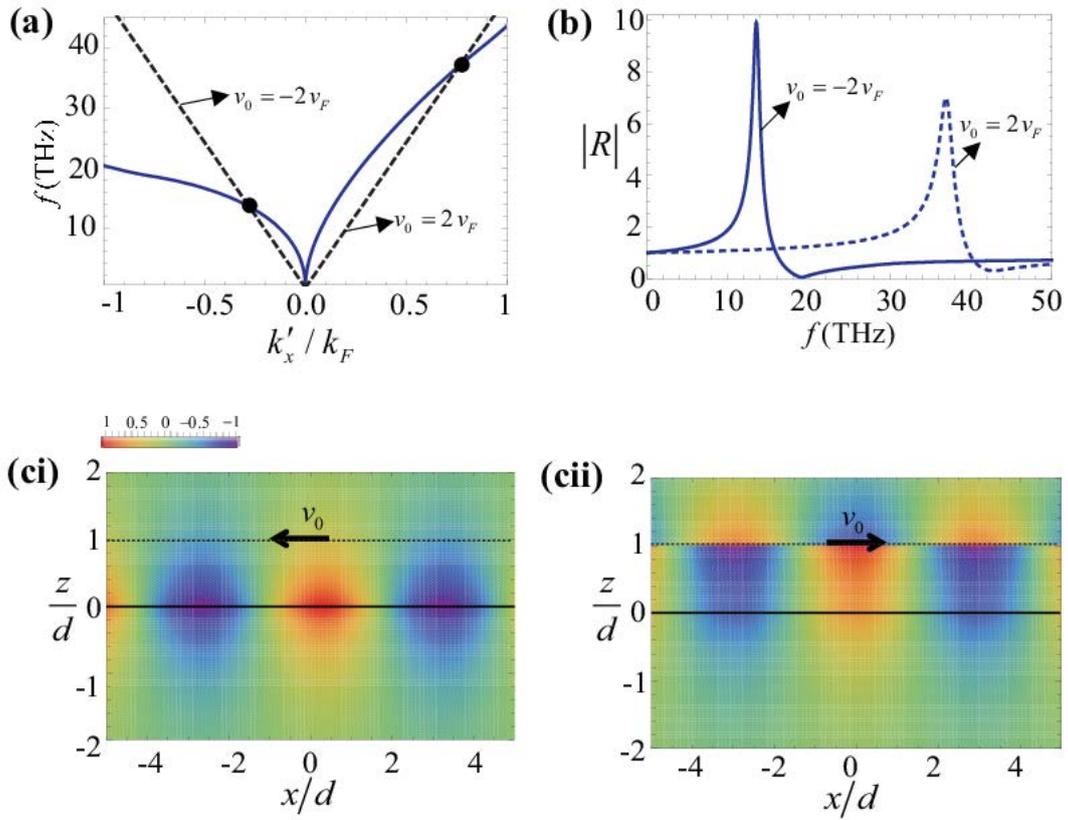

**Fig. 3 (a)** Dispersion diagram of the graphene plasmons for a drift velocity $v_{\text{drift}} = 0.5v_F$ (dark blue curves) superimposed on the lines $\omega = -v_0 k'_x$ and $\omega = v_0 k'_x$, with $v_0 = 2v_F$ (dashed black curves). **(b)** Reflection coefficient $|R|$ as a function of the frequency for the drift velocity $v_{\text{drift}} = 0.5v_F$ and an electron beam with $v_0 = -2v_F$ (dashed curve) or with $v_0 = 2v_F$ (solid curve). **(ci)-(cii)** Density plot of the real part of the normalized magnetic field spectrum $h_y(x, z, \omega)$ calculated at $\omega/2\pi = 13.5\,\text{THz}$ for the drift velocity $v_{\text{drift}} = 0.5v_F$ and $d = 25\,\text{nm}$. **(ci)** The electron beam propagates along $-x$ direction ($v_0 = -2v_F$). **(cii)** The electron beam propagates along $+x$ direction ($v_0 = +2v_F$).

Figures 3ci and 3cii depict a density plot of the spectrum of the radiated magnetic field ($h_y(x,z,\omega)$) evaluated at $\omega/2\pi = 13.5\,\text{THz}$ for $v_0 = \mp 2v_F$, respectively. Consistent with Fig. 3a, for $v_0 = -2v_F$ the emitted field is strongly attached to the graphene sheet ($z = 0$), which is a clear fingerprint of plasmons excitation at $13.5\,\text{THz}$. On the other hand, for $v_0 = +2v_F$ the emitted field at $13.5\,\text{THz}$ is mostly concentrated near the electron beam ($z = d$).

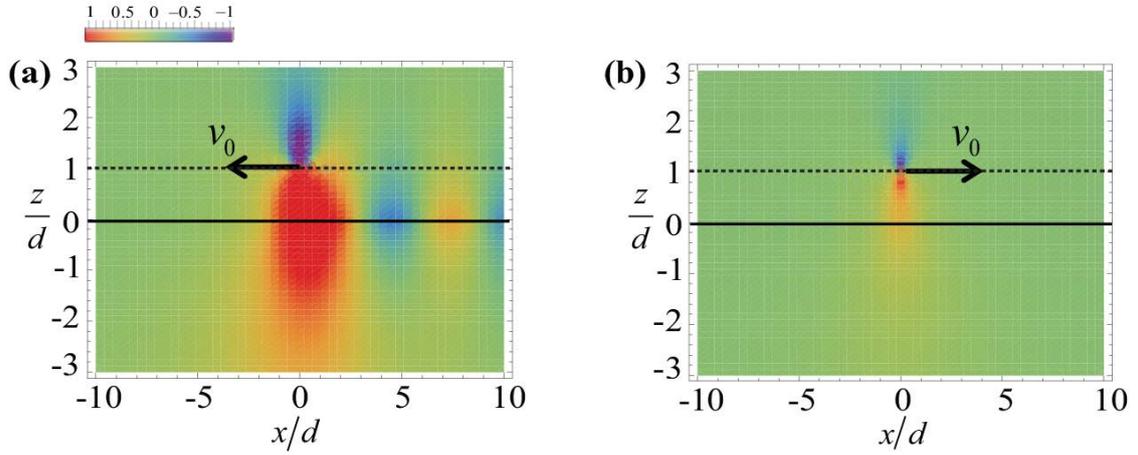

**Fig. 4** Snapshot ($t=0$) of the magnetic field intensity $H_y(x,z,t)$ (in arbitrary units) for a drift velocity $v_{\text{drift}} = 0.5v_F$. The distance between the electron beam and the graphene sheet is $d = 25\,\text{nm}$. The electron beam velocity is **(a)** $v_0 = -2v_F$. **(b)** $v_0 = 2v_F$.

We also calculated the instantaneous field with an inverse Fourier transform [Eq. (1)] for negative and positive values of $v_0$ (see Figs. 4a and 4b, respectively). The results reveal that when the electron beam moves in the direction opposite to the drifting electrons (Fig. 4a) the interaction with the plasmons is stronger due to the longer wavelengths of the excited surface mode (see Fig. 3a). When the beam and moving electrons move along the same direction (Fig. 4b), the excited plasmons have a very short wavelength and die out quickly. Evidently, the emitted radiation field is highly asymmetric and depends strongly on the sign of $v_0$.

## III. Electron energy loss

The amount of energy extracted from the electron beam can be characterized through the stopping power [29, 44, 49] $P_{ext} = -\int \mathbf{E}_{loc} \cdot \mathbf{j}\, dV$, where $\mathbf{E}_{loc}$ represents the local electric field that acts on the electron beam and $\mathbf{j}(x,z,t)$ is the electric current density. Straightforward calculations show that

$$\frac{P_{ext}}{L_y} = \frac{P_0}{z_N} \int_{-\infty}^{+\infty} G(\omega) \frac{d\omega}{\omega_F}, \tag{7}$$

where $P_0 = \dfrac{n_y^2 e^2}{4\pi\varepsilon_0} c$ is a normalization factor (with units of power), $L_y$ represents the width of the current pencil along the $y$ direction, $z_N$ is a reference distance (which we take $z_N = 10\,\text{nm}$) and

$$G(\omega) = \text{Re}\left\{-i\frac{|v_0|}{c}\gamma_0 z_N \frac{\omega_F}{\omega} R\, e^{-2\gamma_0 z_0}\right\} \tag{8}$$

is the (bilateral) dimensionless power spectral density of emitted radiation. The total energy loss after the electron beam travels a distance $L_x$ over the graphene sheet is $\Delta E = P_{ext} L_x / |v_0|$. It can be written as [36]:

$$\Delta E = \int_0^{+\infty} \hbar\omega\, \Gamma_{EELS}(\omega)\, d\omega, \tag{9}$$

where $\Gamma_{EELS}$ is the spectrum of the electron beam loss energy ($\mathcal{E} = \hbar\omega$). It has the explicit formula $\Gamma_{EELS} = \dfrac{A_0}{\omega_F}\Gamma_L$ where $A_0 = \dfrac{n_y^2 e^2}{2\pi\varepsilon_0} \dfrac{1}{\hbar\omega_F} \dfrac{L_x L_y}{z_N}$ is a dimensionless parameter and $\Gamma_L(\omega) = \mathrm{Re}\left\{-i\gamma_0 z_N \left(\dfrac{\omega_F}{\omega}\right)^2 R\, e^{-2\gamma_0 z_0}\right\}$ is a normalized (dimensionless) spectral distribution of the energy loss.

Figure 5a depicts the power spectral density $G$ for the electron beam velocities $v_0 = \pm 2v_F$. Clearly, the spectrum is strongly asymmetric and is peaked near the frequencies that satisfy the selection rule $k_x = \omega/v_0$, analogous to Fig. 3.

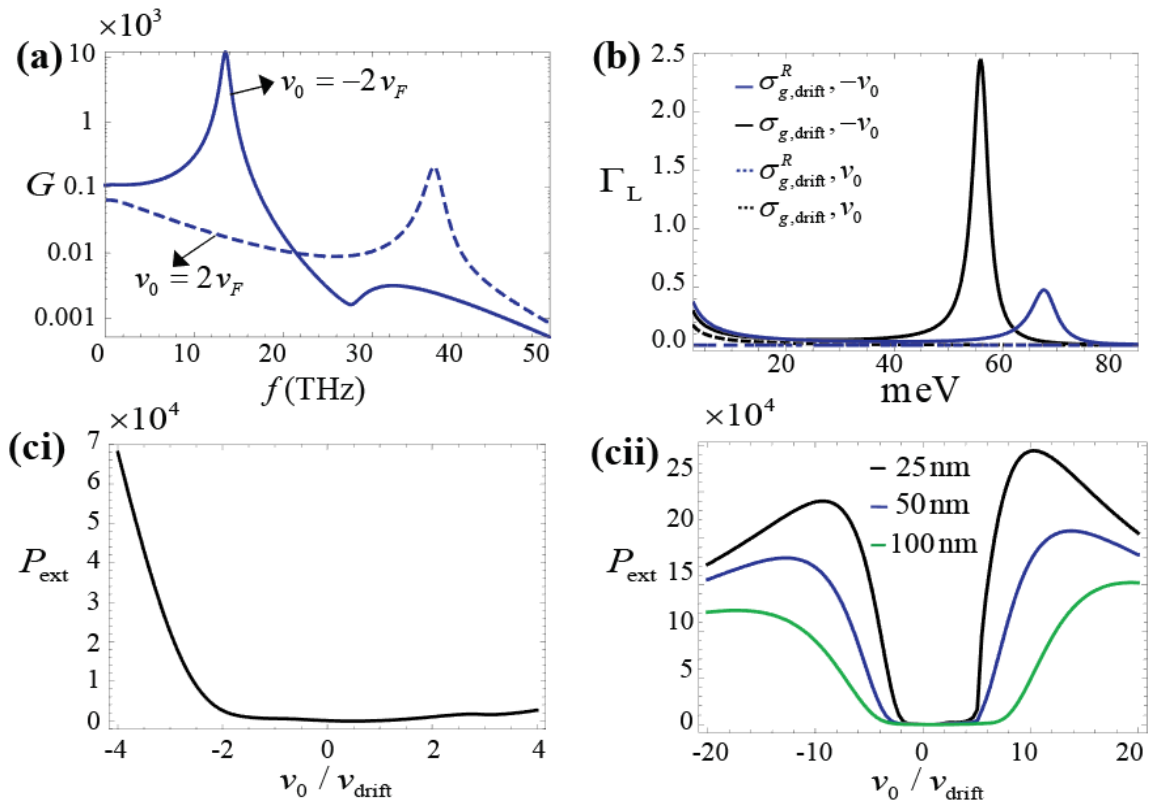

**Fig. 5** (a) Power spectral density $G$ as a function of the frequency, for $v_{\mathrm{drift}} = 0.5 v_F$, $d = 25\,\mathrm{nm}$ and i) $v_0 = -2v_F$ (solid curve) and ii) $v_0 = 2v_F$ (dashed curve). (b) Normalized electron energy loss spectrum calculated with the Galilean (solid black curve) and relativistic (dashed blue curve) models of the dynamic conductivity for $\pm v_0 = 2v_F$. The structural parameters are the same as in panel (a). (ci) Stopping power $P_{\mathrm{ext}}$ (in arbitrary units) for an electron beam with $v_0$ on the order of $v_{\mathrm{drift}}$ and $d = 25\,\mathrm{nm}$. (cii) Expanded representation of the stopping power $P_{\mathrm{ext}}$ as a function of the electron beam velocity $v_0$ for different values of $d$ indicated in the insets.

The normalized spectrum of the energy loss is represented in Fig. 5b as a function of the electron energy ($\mathcal{E} = \hbar\omega$). The black curves are computed using the conductivity obtained with the Galilean transformation [Eq. (4)], whereas the blue curves are calculated using the conductivity obtained with the relativistic Lorentz transformation [Eq. (5)]. The electron energy loss is negligible when the drifting electrons and the electron beam move along the same direction (dotted curves in Fig. 5b). In contrast, when the velocities $v_0$ and $v_{\text{drift}}$ have different signs (solid curves in Fig. 5b) the energy loss spectrum exhibits a resonant peak. As seen in Fig. 5b, the peaks of $\Gamma_L$ predicted by the two conductivity theoretical models do not match and can be shifted by as much as 22 meV. The reason is the relatively weak sensitivity of the graphene plasmons calculated with the relativistic conductivity model to the effect of the drift current bias (see Fig. 2). In principle, low-energy electron microscopy may be able to distinguish such shifts, and thus EELS measurements may instrumental to test the validity of the available theoretical models for the conductivity of graphene with a drift-current bias.

Figure 5ci shows the stopping power for an electron beam with a velocity $v_0$ on the order of $v_{\text{drift}}$. Similar to $\Gamma_L$, the stopping power is negligible when $v_0$ has the same sign as $v_{\text{drift}}$. This result can be explained noting that when the electron beam moves at the same speed as the drifting electrons their interaction is static-like and hence does not lead to any emission of radiation. In contrast, when the velocities have opposite signs the electron beam can exchange energy with the drifting electrons in an irreversible manner due to the relative motion, leading to a strong emission of (Cherenkov-type) radiation. This loss mechanism is reminiscent of (quantum) noncontact electromagnetic friction [30, 54-57]. Note that for the example of Fig. 5ci $|v_0| = 4v_{\text{drift}}$; for this case the stopping power for a negative beam velocity is much larger (about 34 times) than the stopping power for a positive beam velocity. Hence, the drift-current biased graphene can behave similar to "diode" from the point of view of the electron energy loss. For values of $v_0$ much larger than $v_{drift}$, other radiation channels (different from the graphene plasmons) can predominate, and in general the stopping power is larger when the electron beam and the drifting electrons move along the same direction (see Fig. 5cii).

Finally, we study the impact of the nonlocal effects of the bare graphene conductivity on the energy loss spectrum. Figure 6a compares the spectrum of the energy loss $\Gamma_L$ calculated using Eq. (4) with the bare graphene conductivity $\sigma_g(\omega, k_x)$ given by Eq. (6) (nonlocal model) with the result obtained by neglecting the spatial dispersion in the bare-graphene conductivity $\sigma_g \to \sigma_g(\omega, q = 0^+)$ (local model). As before, the energy loss is insignificant when $v_0$ and $v_{drift}$ have the same sign. In contrast, when $v_0$ and $v_{drift}$ have opposite signs, the peak value of $\Gamma_L$ is notoriously sensitive to the nonlocal response of the bare graphene response.

It is difficult to determine how a finite temperature affects $\Gamma_L$ using the nonlocal model of $\sigma_g(\omega, q)$ as its numerical evaluation is rather intricate. To circumvent this problem and have some qualitative understanding on how $\Gamma_L$ may change with the temperature, we calculated the energy loss using the local model with $\sigma_g(\omega)$ evaluated with the standard local (Kubo) formula [32]. As seen in Fig. 6b, a finite temperature weakens somewhat the light-matter interactions, but affects little the position of the peak of $\Gamma_L$.

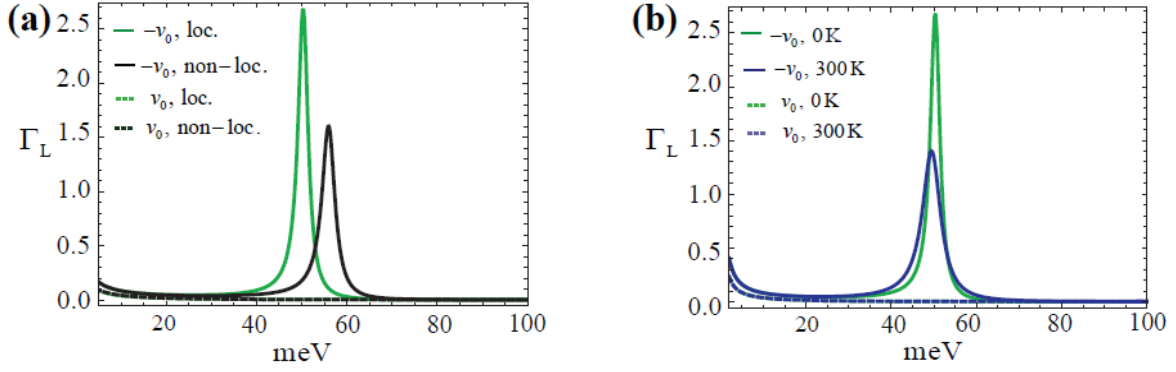

**Fig. 6** Normalized electron energy loss spectrum $\Gamma_L$ for $v_{drift} = 0.5 v_F$, $d = 25\,\mathrm{nm}$ and $v_0 = \pm 2 v_F$, calculated with **(a)** the local and non-local models of $\sigma_g$. **(b)** the local model for T=0K and T=300K.

## IV. Summary

In summary, it was shown that a graphene sheet biased with a drift electric current enables strongly asymmetric light matter interactions that lead to an electron energy loss spectrum strongly dependent on the sign of the electron beam velocity. In particular, the proposed platform may behave as some sort of "Cherenkov-diode", such that the energy loss is strongly suppressed when the drift and electron beam velocities have the same sign and comparable values. When the drift and electron beam velocities have opposite signs, the light-matter interactions are boosted and result in a massive energy loss and in a noncontact friction-type effect. The role of unidirectional graphene plasmons in these effects was highlighted. It was suggested that EELS measurements may be helpful to clarify the validity of the available theoretical conductivity models of graphene with a drift current bias, as the peaks of energy loss predicted by the different models can be shifted by as much as 20meV.

# Appendix A

The function $G(z)$ in Eq. (6c) is given by:

$$G(z) = z\,\mathrm{sq}(z-1)\mathrm{sq}(z+1) - \left[\ln\left((z+\mathrm{sq}(z-1)\mathrm{sq}(z+1))e^{i\theta_0}\right) - i\theta_0\right] \quad (A1)$$

where $\ln$ represents the standard logarithm function with a branch cut in the negative real axis and $\theta_0 = -\pi/4$. The function $\mathrm{sq}(w)$ with $w = w' + iw''$ is determined by

$$\mathrm{sq}(w) = \begin{cases} -\sqrt{w}, & \text{if } w' < 0 \text{ and } w'' < 0 \\ \sqrt{w}, & \text{otherwise} \end{cases} \quad (A2)$$

where $\sqrt{\phantom{w}}$ is the standard square root function with a branch cut in the negative real axis.

**Acknowledgements:**

This work is supported in part by the IET under the A F Harvey Engineering Research Prize 2018 and by Fundação para Ciência e a Tecnologia (FCT) under project PTDC/EEITEL/4543/2014 and UID/EEA/50008/2019. F. R. Prudêncio acknowledges financial support by FCT under the Post-Doctoral fellowship SFRH/BPD/108823/2015.